# On the Magnetic Excitation Spectra of High $T_c$ Cu Oxides up to the Energies far above the Resonance Energy

Satoshi Iikubo, Masafumi Ito, Akito Kobayashi, Masatoshi Sato*

*Department of Physics, Division of Material Science, Nagoya University,*

Furo-cho, Chikusa-ku, Nagoya 464-8602

and

Kazuhisa Kakurai

Advanced Science Research center, JAERI, Tokai, Ibaraki 319-1195

Magnetic excitation spectra $\chi''(q,\omega)$ of $YBa_2Cu_3O_y$ and La214 systems have been studied. For $La_{1.88}Sr_{0.12}CuO_4$, $\chi''(q,\omega)$ have been measured up to ~30 meV and existing data have been analyzed up to the energy $\omega$~150 meV by using the phenomenological expression of the generalized magnetic susceptibility $\chi(q,\omega)=\chi_0(q,\omega)/\{1+J(q)\chi_0(q,\omega)\}$, where $\chi_0(q,\omega)$ is the susceptibility of the electrons without the exchange coupling $J(q)$ among them. In the relatively low energy region up to slightly above the resonance energy $\omega_r$, it has been reported by the authors' group that the expression can explain characteristics of the $q$- and $\omega$-dependence of the spectra of $YBa_2Cu_3O_y$ (YBCO or $YBCO_y$). Here, it is also pointed out that the expression can reproduce the rotation of four incommensurate peaks of $\chi''(q,\omega)$ within the $a^*$-$b^*$ plane about ($\pi/a$, $\pi/a$) {or so-called ($\pi$, $\pi$)} point by 45 °, which occurs as $\omega$ goes to the energy region far above $\omega_r$ from $\omega$ below $\omega_r$. For $La_{2-x}Sr_xCuO_4$ and $La_{2-x}Ba_xCuO_4$, agreements between the observed results and the calculations are less satisfactory than for YBCO, indicating that we have to take account of the existence of the "stripes" to consistently explain the observed $\chi''(q,\omega)$ of La214 system especially near $x$=1/8.

Keywords: magnetic excitation spectra, neutron scattering, stripes

* Corresponding author. E-mail address: e43247a@nucc.cc.nagoya-u.ac.jp

In Cu oxides, the active magnetism is closely connected with the occurrence of the high-$T_c$ superconductivity and the anomalous normal state.[1-4] Neutron scattering is a powerful method to investigate magnetic characteristics of the systems and extensive studies by this means have revealed various important characteristics of magnetic behaviors of high $T_c$ systems. The pseudo gap behavior observed in the magnetic excitation spectra $\chi''(\boldsymbol{q},\omega)$ of $YBa_2Cu_3O_y$ (YBCO or $YBCO_y$) even far above $T_c$ [5,6] is one of examples of such characteristics. Now, the behaviors related to the “stripe” ordering or “ stripe” fluctuations are being studied quite extensively.

In arguing the spectra $\chi''(\boldsymbol{q},\omega)$, the expression

$$\chi(\boldsymbol{q},\omega) =\chi_0(\boldsymbol{q},\omega) /\{1+J(\boldsymbol{q})\chi_0(\boldsymbol{q},\omega) \}, \quad (1)$$

is often used [7,8], where $\chi_0(\boldsymbol{q},\omega)$ is the generalized susceptibility of the strongly correlated electrons without the exchange coupling $J(\boldsymbol{q})$ ($=J_0(\cos q_x a+\cos q_y a)$ with $a$ being the lattice constant). We have been studying our own results and other reported data of $\chi''(\boldsymbol{q},\omega)$ by using the expression,[9-11] where various experimental informations, the incommensurate(IC) –commensurate (so-called resonance peak) -IC variation of $\chi''(\boldsymbol{q},\omega)$ with increasing ω first observed for YBCO[13-15] and differences between detailed $\boldsymbol{q}$- and ω-dependences of the observed spectra of YBCO and La214 systems such as $La_{2-x-y}Nd_ySr_xCuO_4$ (LNSCO) and $La_{2-x}Sr_xCuO_4$ (LSCO), have been used in choosing realistic material parameters. *Effective* band parameters $t_0$, $t_1$ and $t_2$, the ratios of which are chosen to reproduce the Fermi surface shape ($t_0$:$t_1$:$t_2$=1:-1/6:1/5 for $YBa_2Cu_3O_y$[16] and 1:-1/6:0[16] for La214 systems) have been used. The energy gap is expressed as $\Delta_s(k)=(\Delta_0/2)(\cos k_x a-\cos k_y a)$.[17] (The gap is assumed to be $T$-independent in the underdoped region of YBCO by considering the existence of the pseudo gap.) Further details of the calculations can be found in refs. 9-12 and 17.

The results of these studies can be summarized as follows. For YBCO, the calculations can well reproduce the data of $\chi''(\boldsymbol{q},\omega)$, if we consider the quasi particle broadening $\Gamma(\varepsilon)$, which has drastic $T$- and energy(ε)-dependence in the region of ε smaller than the maximum gap $2\Delta_0$, as observed by other kinds of experimental studies. The introduction of large Γ and $\Delta_0$ has been found to suppress the magnetic ordering. On the other hand, we have to use very small Γ values to reproduce the sharp width of the IC peaks observed for La214 systems. Because the Γ value we have to use is the smallest at $x\sim 1/8$, we attribute it to possible effects of the “stripe” order or “stripe” fluctuations. The clear contrast between the results for YBCO and La214 systems suggests that the “stripes” and their fluctuations are important only in the latter systems.

Recently, the magnetic neutron scattering intensities $S(\boldsymbol{q},\omega)$, which is defined as $(n+1)\times \chi''(\boldsymbol{q},\omega)$ with $n$ being the Bose factor and can be approximated by $\chi''(\boldsymbol{q},\omega)$ for $\omega >> k_B T$, have been reported for $YBCO_{6.6}$ and $La_{1.875}Ba_{0.125}CuO_4$ (LBCO or $LBCO_{0.125}$) by Hayden *et al.*[18] and Tranquada *et al.*,[19] respectively, in the two dimensional ($q_x$-$q_y$) reciprocal space up to the energy far above their resonance energy $\omega_r$. Both systems exhibit the IC- commensurate (C)-IC variation with ω. The rotation of the IC peak positions by 45 ° within the $\boldsymbol{a}^*$-$\boldsymbol{b}^*$ plane about ($\pi/a$, $\pi/a$) {or so-called (π, π)} point is clearly observed in YBCO and less clearly in LBCO, when the energy ω is increased from the low energy region ($\omega<\omega_r$ ) to the region far above $\omega_r$. Because in our previous works, we did not study the $E$ region far above $\omega_r$, it is now interesting to study here if the rotation can be explained by the above expression or the "stripe" correlation is required for the explanation of the rotation.

We have extended the calculation to the higher energy region in the two-dimensional $\boldsymbol{q}$ space and found that the calculation can reproduce four incommensurate peaks at the diagonal IC $\boldsymbol{q}$ points (1/2+δ',1/2+δ'), (1/2−δ',1/2−δ'), (1/2+δ',1/2−δ') and (1/2−δ',1/2+δ') in the energy region far above $\omega_r$, in agreement with the experimental results for YBCO.

For La214 system, $\chi''(\boldsymbol{q},\omega)$ of $La_{1.88}Sr_{0.12}CuO_4$ have been newly taken and analyzed to see effects of "stripes" on the excitation spectra. We have found that the incommensurate peaks of $La_{1.88}Sr_{0.12}CuO_4$ in the low energy region are too sharp to be reproduced by the calculations, as was found previously for $La_{1.48}Nd_{0.4}Sr_{0.12}CuO_4$.[11] We calculate $\chi''(\boldsymbol{q},\omega)$ up to the very large value of ω far above $E_r$ and compare the results with the existing data.

Here, we calculate $\chi''(\boldsymbol{q},\omega)$ in the way essentially similar to the previous ones.[9-12] The exchange coupling $J_0$ is so chosen that the resonance peak energy observed in $\chi''(\boldsymbol{q},\omega)$ can reasonably be reproduced. The intra atomic Coulomb interaction $U$ is set to be zero. (We just consider the strong correlation effect by adopting a small effective values of the band parameters which gives the band width similar to that obtained by the *d-p* model for the coherent in-gap band[17] or the holon band of the *t-J* model.[8, 16])

For YBCO, the value of $2|\Delta_0|$ is assumed, as in the previous calculations, to be $T$ independent in the underdoped region by considering the persistence of the pseudo gap up to very high temperature. Figure 1 shows the simple model of the Γ-ε curves used for YBCO,[9,10,12] where $\Gamma_0$ and $\Gamma_h$ are the values at the chemical potential μ and at $|\varepsilon|>2\Delta_0$, respectively. $\Gamma_h$ is assumed to be $T$ independent, while $\Gamma_0$ increases with increasing $T$. For $LBCO_{0.125}$, we use the curve shown in the inset of Fig. 1, where we set the superconducting gap to be zero.

For both systems, the μ value is determined to adjust the $\boldsymbol{q}$-position or the incommensurability δ of the observed peaks of $\chi''(\boldsymbol{q},\omega)$ at low temperatures and low energies. The numerical calculations of $\chi''(\boldsymbol{q},\omega)$ have been carried out as described in refs. 9-12, where the steps of the integrations over $\boldsymbol{q}$ and ω are fine enough to obtain reliable results of the integrations.

In the neutron scattering measurements on $La_{1.88}Sr_{0.12}CuO_4$, five aligned crystals with the typical size of about 6 mmϕ ×30 mm were used. The triple axis spectrometer ISSP-PONTA installed at JRR-3M of JAERI in Tokai was used. The horizontal collimations were 40'-40'-80'-80'. The sample crystals were oriented with the [001] axis vertical.

Figure 2 shows the intensity map of the magnetic scattering in the two-dimensional reciprocal space calculated for $YBCO_{6.6}$ at 10 K with the neutron transfer energies ω=24, 34, 65, 75 and 85 meV. Used parameters are $t_0$=-20 meV, $2\Delta_0$=100 meV, $J_0$=72 meV and μ =-8 meV. In the low energy region, the peaks of $\chi''(\boldsymbol{q},\omega)$ are located at the $\boldsymbol{q}$ points (1/2, 1/2±δ) and (1/2±δ, 1/2) Interestingly, at the energies ~85 meV(> $\omega_r$=34 meV), we can see peaks at the $\boldsymbol{q}$ points (1/2+δ', 1/2+δ'), (1/2−δ', 1/2−δ'), (1/2+δ', 1/2−δ') and (1/2−δ', 1/2+δ'), that is, the directions towards these peak points from ($\pi/a$, $\pi/a$ ) are shifted by 45 ° from those observed at $\omega<\omega_r$, which is in good agreement with the experimental observation by Hayden *et al.*[18] The map reproduces the observed distribution of $\chi''(\boldsymbol{q},\omega)$ reasonably well. Stock *et al.* [22] have shown the ring-like distribution of the spectral intensity at ω=77. 5 meV at *T*=6 K, which is also similar to the result of the present calculation at ω=75 meV. According to our calculations, this type of diagonal IC peaks can be found for relatively large values of the superconducting gap $2\Delta_0$ ($2\Delta_0$>60 meV). Figure 3 shows the *T*-dependence of the intensity map of the magnetic scattering calculated for $YBCO_{6.6}$ at ω=85 meV. In the calculation, we have used the *T*-dependence of $\Gamma_0$ shown in Fig. 1 ($\Gamma_0$=4, 10 and 60 meV at *T*=10, 66 and 300 K, respectively).[10,12] The diagonal incommensurate peaks collapse to a broad single peak as *T* is increased. Here, it should be noted, however, that the calculated value of δ' at ω~85 meV is larger than the observed one. We think that the model, which uses the effective band parameters, is too simple to describe the observed spectra well over the very wide ω region.

We have also studied to what extent the data of La214 system can be reproduced by the calculation based on eq. (1). In Figs. 4(a) and 4(b), the scattering intensity $S(\boldsymbol{q},\omega)$ {$\equiv(n+1)\chi''(\boldsymbol{q},\omega)$, *n* being the Bose factor} of $La_{1.88}Sr_{0.12}CuO_4$ taken at *T*=38 K for various values of ω, and $S(\boldsymbol{q},\omega)$ taken for various values of *T* at ω=2.5 meV are shown, respectively. In the figures, the curves fitted by using eq. (1) are also shown, where the following parameters are used. $t_0$=-40 meV, $2\Delta_0$=0, $J_0$=45 meV, μ=-35 meV and $\Gamma_0=\Gamma_h$=10 meV. We find that due to the significant “stripe” correlation or fluctuations, the profiles at relatively small ω and low temperatures are too sharp to be reproduced by the calculations, which are consistent with our previous results.[11,12] Using this parameter set used for $La_{1.88}Sr_{0.12}CuO_4$, we have calculated $\chi''(\boldsymbol{q},\omega)$ at 12 K up to very high energies, and found that although the IC-C-IC variation with ω is reproduced, weak but discernible IC peaks in the map of the calculated $\chi''(\boldsymbol{q},\omega)$ far above $E_r$ are at points (1/2, 1/2±δ”) and (1/2±δ”, 1/2). It seems not to be consistent with the experimentally observed data of $LBCO_{0.125}$.[19] To reproduce the ring-like structure of the $\chi''(\boldsymbol{q},\omega)$ map far above $\omega_r$, we have to use the $|t_0|$ value smaller than 30 meV. Figure 5 shows the map for $t_0$=−25 meV and μ~−25 meV. At ω= 155 meV, the distribution of the intensity seems to be ring-like, which is roughly consistent with the experimental result[19] for $LBCO_{0.125}$ though in this energy range, weak diagonal IC peaks may be distinguishable in the observed maps. (Strictly speaking, the excitation spectra $\chi''(\boldsymbol{q},\omega)$ of $LBCO_{0.125}$ and $La_{1.88}Sr_{0.12}CuO_4$ are different from each other's, because only the former has the clear “stripe” ordering, and the parameters for these systems may have to be different.)

The present results indicate that for the parameters which are chosen to give the best fit to the observed $\chi''(\boldsymbol{q},\omega)$ data of $La_{1.88}Sr_{0.12}CuO_4$ in the low energy region (note that the fitting is not good enough even though it is the best, because due to the “stripe” correlation the IC peaks in the low energy region is too sharp.), do not completely reproduce the high energy map. Instead, a slightly different values of $t_0$ and μ roughly reproduce the features of the observed map in the energy region far above $\omega_r$. We think that the present model, which uses effective band parameters is too simple, as in the case of YBCO, to describe the very wide region of $\omega$.

We have so far shown the results of the model calculations comparing with the experimental observations. For YBCO, the model calculations can reproduce the qualitative characteristics of the ***q***- and $\omega$-dependences of $\chi''(\boldsymbol{q},\omega)$. We just emphasize here that the calculation can reproduce the observed rotation of the four peaks of $\chi''(\boldsymbol{q},\omega)$ by 45 ° within the ***a*-b**** plane about ($\pi/a$, $\pi/a$), when $\omega$ increases to ~85 meV from the low energy region $<\omega_r$. Then, the rotation in this calculation is not due to the existence of the "stripe" correlation. Norman *et al.*[23] have reported the possible existence of the diagonal IC peaks above $E_r$ by using the RPA form of $\chi(\boldsymbol{q},\omega)$ similar to eq. (1). However, They did not predict the observed rotation of the IC peak positions with increasing $\omega$ through $\omega_r$ for a given set of parameters. Recently, Hinkov *et al.*[24] have presented data that $\chi''(\boldsymbol{q},\omega)$ of YBCO system has two-dimensional nature. It may support, we think, an idea that the magnetic excitation spectra can essentially be understood by the itinerant electron picture, though the observed $\omega$-dependent anisotropy of the spectra $\chi''(\boldsymbol{q},\omega)$, cannot simply be explained.

For $La_{1.88}Sr_{0.12}CuO_4$, the observed peak widths in the low energy region are so small that the calculation cannot reproduce the observation even if we use rather small values of $\Gamma_h$. This seems to be understood by introducing the effects of the slowly fluctuating "stripes".

The results of the present calculations seem not to be very different from those of the calculations carried out by considering "stripes".[20,21] Rough features of $\chi''(\boldsymbol{q},\omega)$ such as the IC-C-IC variation with $\omega$ can be obtained above the "stripe" ordering temperature (if it exists) irrespective of whether the "stripe" correlation is considered or not especially in the high energy region. (Below $T_e$, the two-dimensional nature found in the calculations of refs. 20 and 21 is obtained after averaging the spectra over the two directions of the "stripes".) Of course, details are different between the present report and those of refs. 20 and 21: The latter two predict the clear intensity peaks at the diagonal IC points above $\omega_r$, while the former just show the ring-like distribution.for La214 system. (The diagonal IC peaks in YBCO arise from the existence of the gap $2\Delta_0$. The observed intensity map of $LBCO_{0.125}$ seems to have the almost ring-like structure, though we may see weak diagonal IC peaks superposed on the ring. The superposed peaks are, we think, due to the effect of "stripes".

We think that effects of "stripes" can be seen only in the detailed features of $\chi''(\boldsymbol{q},\omega)$ such as the sharp peak width in the low energy region. On this points, the data of $La_{1.6-x}Nd_{0.4}Sr_xCuO_4$ published by the present authors' group seem to be informative.[11] In $La_{1.6-x}Nd_{0.4}Sr_xCuO_4$, the quasi particle broadening $\Gamma$ is suppressed in the low energy region as $T$ approaches the "stripe" ordering temperature $T_e$ from above or as the "stripe" correlation grows above $T_e$. Probably due to this suppression, the profile widths of the incommensurate peaks of $\chi''(\boldsymbol{q},\omega)$ cannot be reproduced by the above expression even above $T_e$ in La214.

In summary, results of the studies on the magnetic excitation spectra of $YBa_2Cu_3O_y$ and La214 systems have been presented up to very high energies far above $\omega_r$. For the former system, the observed characteristics, including the rotation of the four IC peaks about the so-called (π,π) points, can be explained without considering "stripes", indicating that it cannot simply be considered as an effect of "stripe" correlation. For La214, the very small peak width observed at low energies is the manifestation of the "stripe" correlation.

**Acknowledgements-** The present work is supported by grants-in-Aid for Scientific Research from the Japan Society for the promotion of Science and Grants-in Aid on Priority Areas from the Ministry of Education, Culture, Sports, Science and Technology of Japan. One of the authors (S. I) is supported by a Research Fellowship of the Japan Society for the Promotion of Science for Young Scientists.

Figure captions

Fig. 1 Model functions of $\Gamma(\varepsilon)$ used in the present calculations of $\chi''(\boldsymbol{q},\omega)$ for YBCO are shown at several temperatures. $\Gamma_0$ and $\Gamma_h$ correspond to the values at $\mu=0$ and $|\varepsilon|>2\Delta_0$, respectively. Inset shows the model function of $\Gamma(\varepsilon)$ used for LBCO (see text for details).

Fig. 2 Maps of $\chi''(\boldsymbol{q},\omega)$ calculated for YBCO at 10 K with the neutron transfer energies $\omega=24$, 34, 65, 75 and 85 meV.

Fig. 3 $T$-dependence of $\chi''(\boldsymbol{q},\omega)$ of the magnetic scattering calculated for YBCO at $\omega=85$ meV.

Fig. 4 (a) Magnetic excitation spectra taken at $T$=38 K for $La_{1.88}Sr_{0.12}CuO_4$ by scanning along (0.5,$k$,0) at the fixed energies of $\omega=2.5$, 5 and 8 meV and along (1.5,k,0) at $\omega=11$, 14 and 30 meV, respectively.The intensities of the scans along two different lines are normalized at $\omega=8$ meV. Solid lines show the calculated spectra $\chi''(\boldsymbol{q},\omega)$ obtained for the parameters $t_0=$ 40 meV, $2\Delta_0$~0 meV, $J_0$=45 meV, $\Gamma_. =\Gamma_0=\Gamma_{h.}=10$ meV. (b) The spectral intensities $\chi''(\boldsymbol{q},\omega)$ taken with the transfer energy of $\omega=2.5$ meV are shown at three different temperatures. Solid lines show the calculated spectra $\chi''(\boldsymbol{q},\omega)$ obtained for the same parameters as used in (a).

Fig. 5 (a) Maps of $\chi''(\boldsymbol{q},\omega)$ calculated at 12 K for La214 system with $x$=0.125 or $LBCO_{0.125}$ are shown for $\omega=$ 5, 55 and 155 meV.

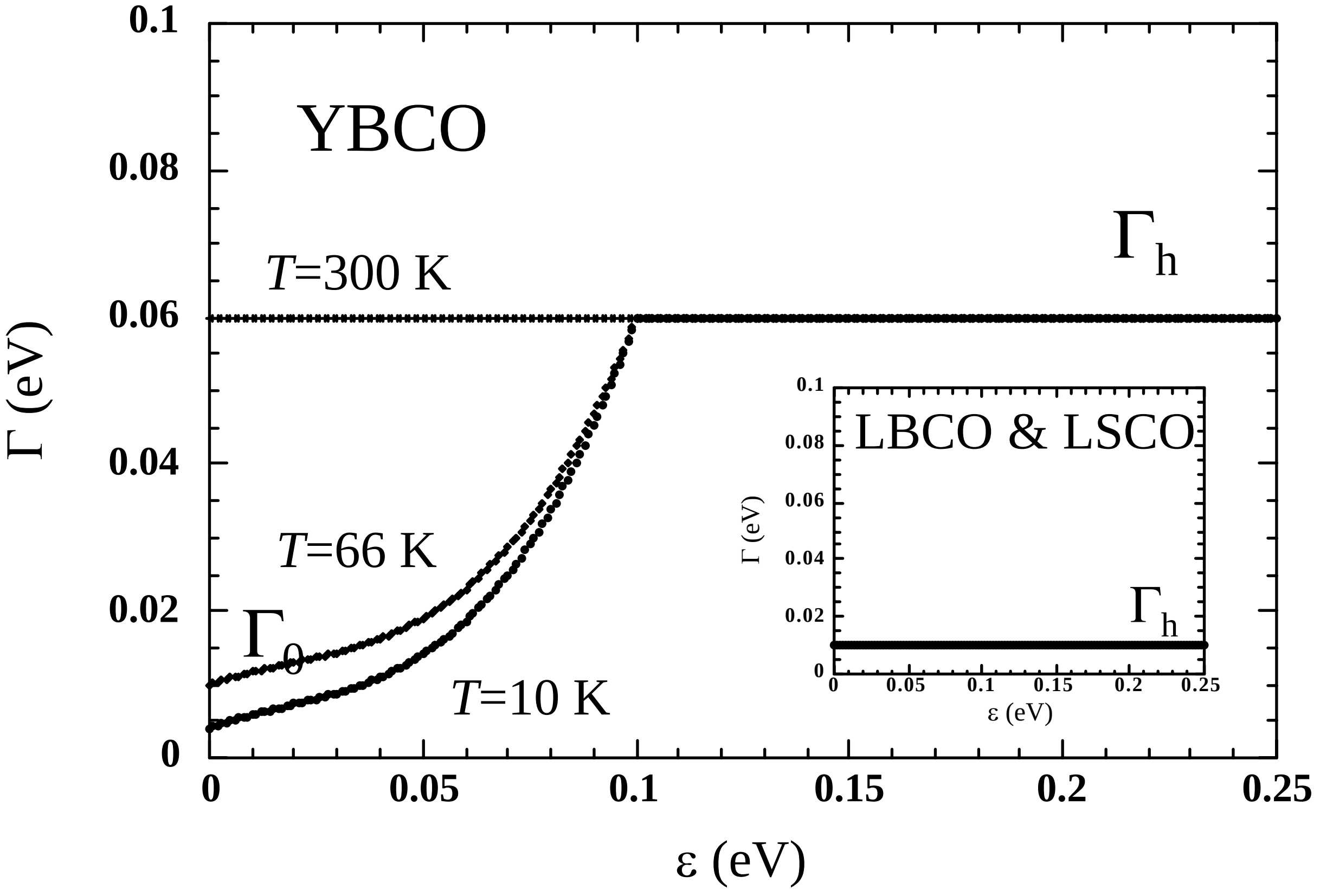

Fig. 1

S. Iikubo *et al.*

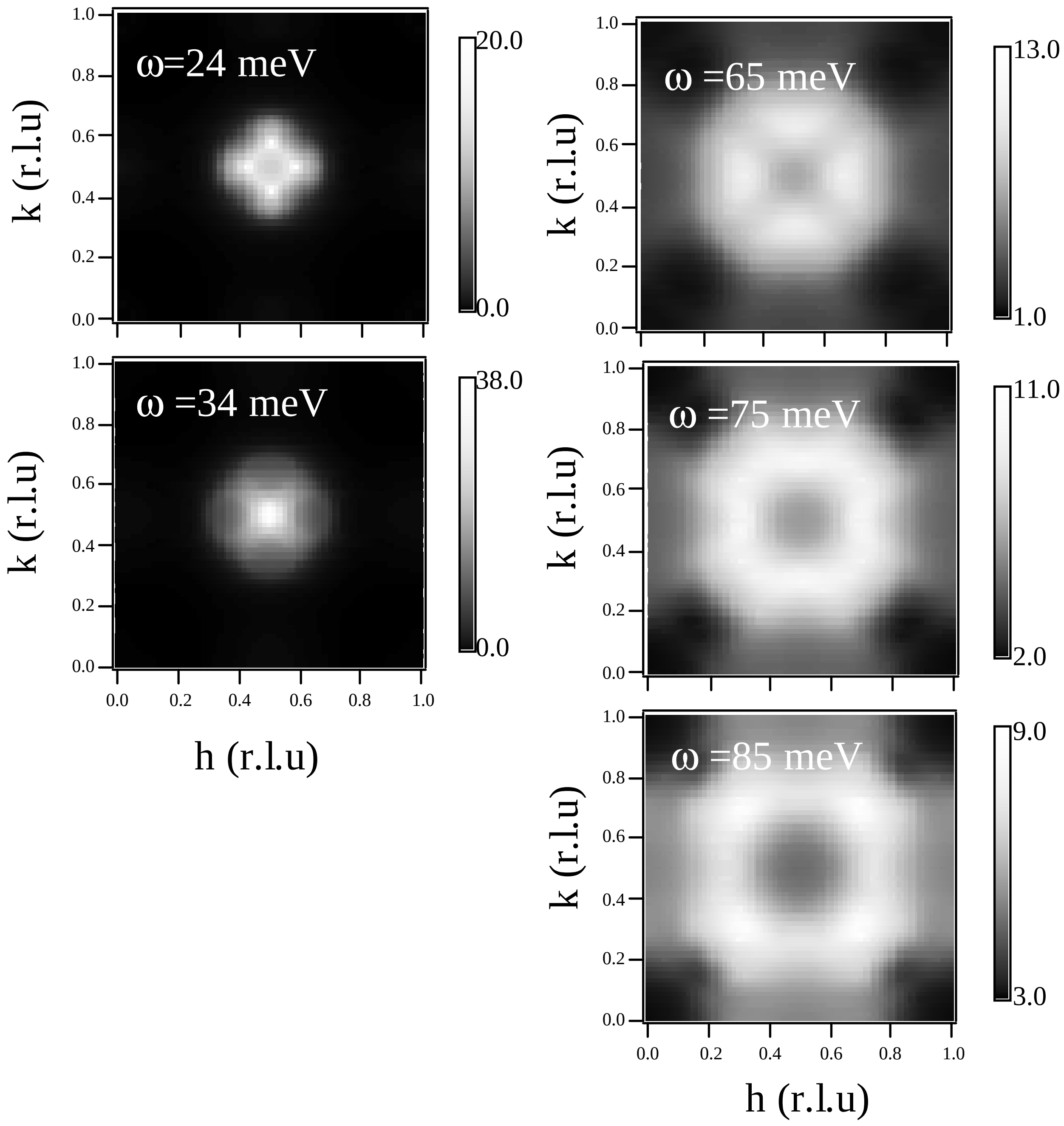

Fig. 2
S. Iikubo *et al.*

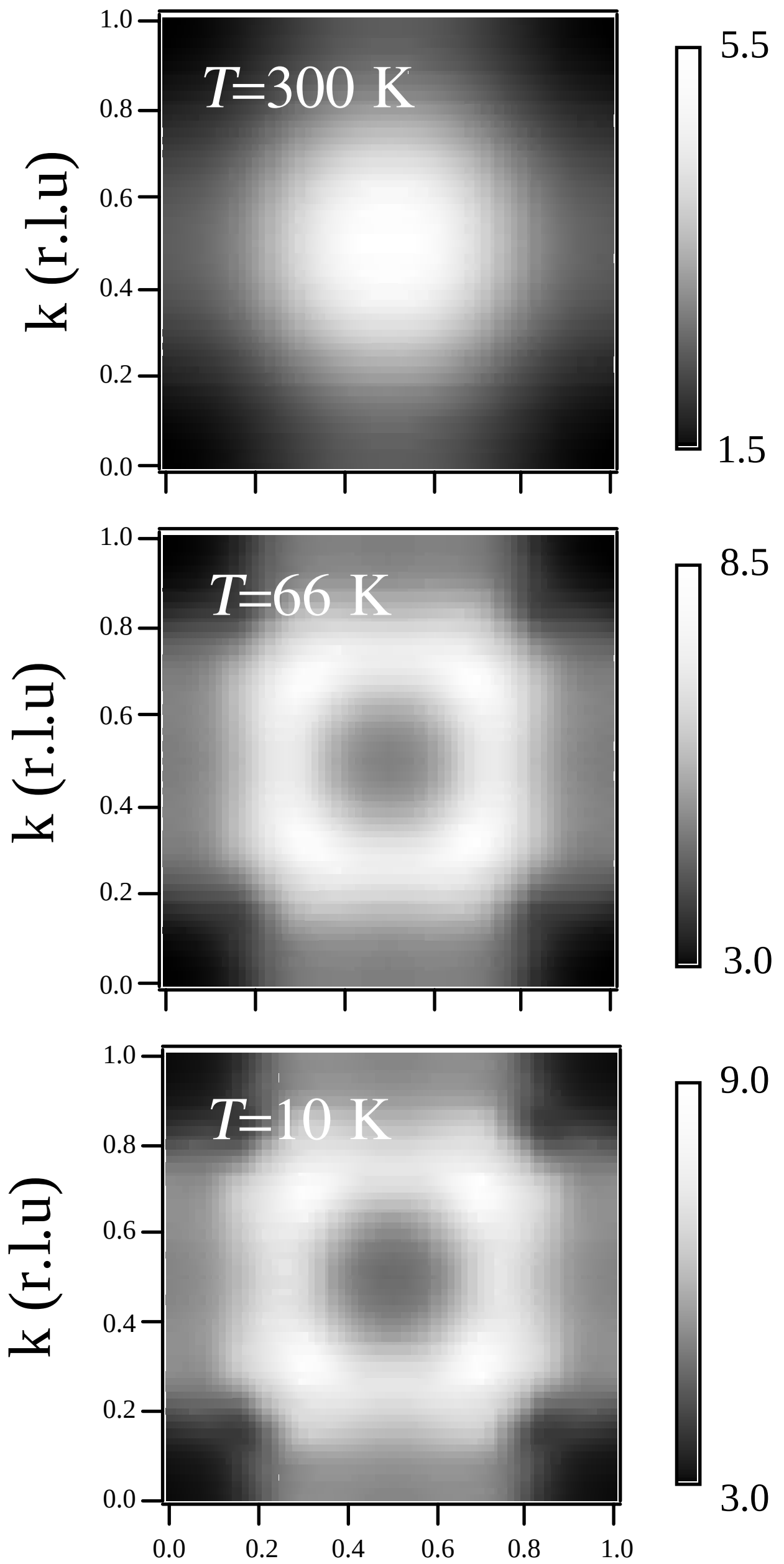

Fig. 3

S. Iikubo *et al.*

# $La_{1.88}Sr_{0.12}CuO_4$

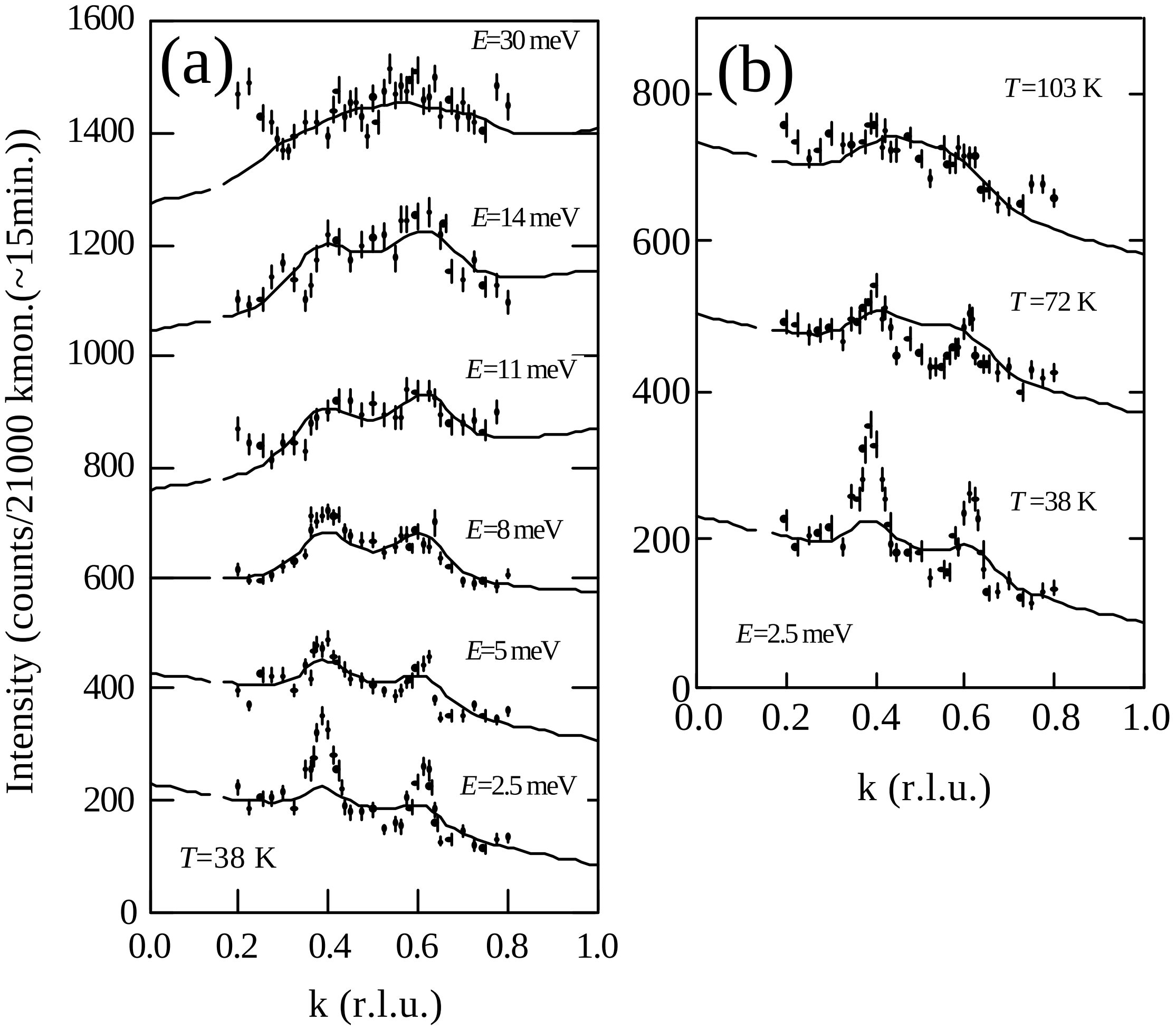

Fig. 4(a), (b)
S. Iikubo *et al.*

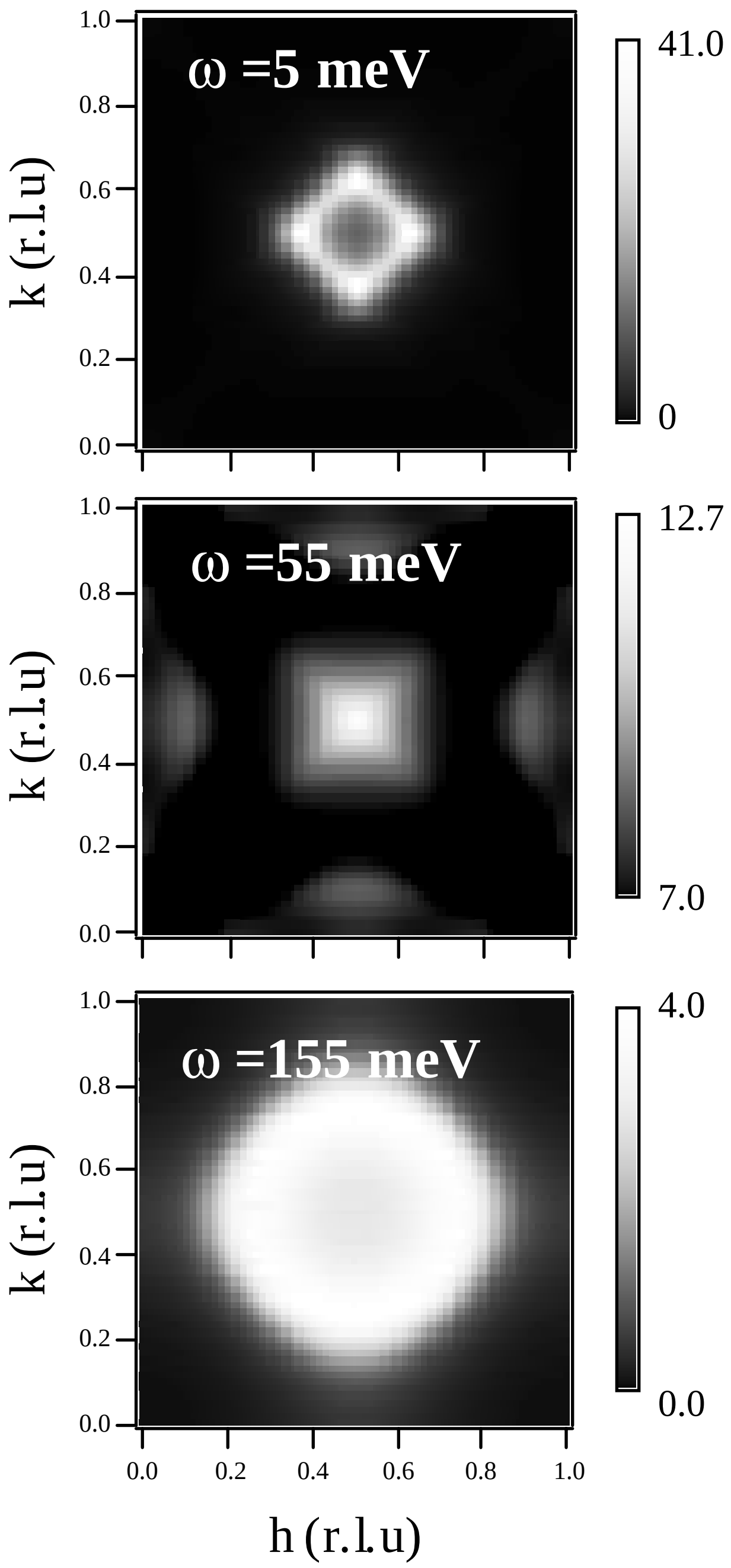

Fig. 5

S. Iikubo *et al.*